# Engineering nonlinear Hall effect in bilayer graphene/black phosphorus heterostructures

Xing-Guo Ye,[1] Zhen-Tao Zhang,[1] Peng-Fei Zhu,[1] Wen-Zheng Xu,[1] An-Qi Wang,[1] and Zhi-Min Liao[1,2,*]

[1]*State Key Laboratory for Mesoscopic Physics and Frontiers Science Center for Nano-optoelectronics,*
*School of Physics, Peking University, Beijing 100871, China*
[2]*Hefei National Laboratory, Hefei 230088, China*

Two-dimensional van der Waals materials offer a highly tunable platform for generating emergent quantum phenomena through symmetry breaking. Stacking-induced symmetry breaking at interfaces provides an effective method to modulate their electronic properties for functional devices. Here, we strategically stack bilayer graphene with black phosphorus, a low-symmetry semiconductor, to break the symmetries and induce the nonlinear Hall effect (NLHE) that can persist up to room temperature. Intriguingly, it is found the NLHE undergoes sign reversals by varying the electrical displacement field under fixed carrier density. The scaling analysis reveals that the sign reversal of the NLHE is contributed from both the Berry curvature dipole (BCD) and extrinsic scatterings. The displacement field-induced sign reversal of the BCD indicates asymmetric distributions of Berry curvature hot spots across different Fermi pockets in bilayer graphene. Our findings suggest that symmetry engineering of van der Waals heterostructures is promising for room-temperature applications based on nonlinear quantum devices, such as high-frequency rectifiers and wireless charging.

*Introduction.* The nonlinear Hall effect (NLHE) has recently attracted significant research interest [1–8]. The NLHE refers to the nonlinear Hall response to an applied current, and its intrinsic origin is thought to lie in the Berry curvature dipole (BCD), an important topological geometric quantity [1,2]. In two-dimensional (2D) crystals, the BCD is defined as $\Lambda_\alpha = -\frac{1}{\hbar} \int \delta(\varepsilon - \varepsilon_F) \frac{\partial \varepsilon}{\partial k_\alpha} \Omega_z d^2\mathbf{k}$, where $\varepsilon$ is the energy, $\varepsilon_F$ is the Fermi level, $k$ is the wave vector, $\frac{\partial \varepsilon}{\partial k_\alpha}$ is the band slope along $\alpha$, and $\Omega_z$ is the Berry curvature [1]. The delta function $\delta(\varepsilon - \varepsilon_F)$ restricts contributions to states near the Fermi surface. The BCD plays a central role in nonlinear quantum phenomena like the NLHE [1,2] and the circular photogalvanic effect [9–11]. Additionally, scattering mechanisms such as skew scattering and side jump may contribute to the NLHE [3]. The NLHE has potential applications in nonlinear quantum devices, including high-frequency rectifiers, wireless charging, and energy harvesting [3].

Generally, the NLHE is permitted under time-reversal symmetry but requires broken inversion symmetry [1]. Experimental observations of the NLHE have thus far been limited to low-symmetry systems such as Weyl semimetals like few-layer WTe$_2$ [12–15] and TaIrTe$_4$ [16], strain-modulated monolayers of MoS$_2$ and WSe$_2$ [17–19], and other select topological materials [20–23]. Interface engineering presents an alternative route to achieving the nonzero NLHE, especially in moiré structures, including twisted WSe$_2$ [24], twisted graphene [25–28], and graphene/hBN superlattices [29]. The NLHE has also been demonstrated in corrugated bilayer graphene via artificial strain [30]. However, room-temperature NLHE observations remain rare [16,31–33], limiting its practical applications. Further exploration of the BCD-induced NLHE at van der Waals heterointerfaces with broken symmetry, along with its dependence on the electric displacement field ($D$), Fermi energy, and temperature, would be valuable for advancing device applications.

In this letter, we demonstrate the generation and modulation of the NLHE through symmetry engineering in bilayer graphene/black phosphorus (BP) van der Waals heterostructures. By stacking bilayer graphene ($C_3$ symmetry) with BP ($C_2$ symmetry) to lower the symmetries, the NLHE is observed at room temperature. Notably, the NLHE undergoes sign reversal with changes in the displacement field at a fixed carrier density, which is attributed to the corresponding sign change in the BCD that is identified through scaling analysis. Numerical calculations link this BCD reversal to effective lateral strain from moiré superlattice formation in the heterostructure. Our findings establish bilayer graphene/BP as a tunable, room-temperature platform for nonlinear devices by engineering the Fermi surface topology.

*NLHE due to symmetry breaking.* Bilayer graphene/BP devices are encapsulated by hBN layers, with top and bottom gates, $V_T$ and $V_B$, as illustrated in Fig. 1(a). This dual gate configuration allows independent control of carrier density $n$ and $D$ (Refs. [34–40]). An optical image of the device is shown in Fig. 1(b), showing multielectrodes in a circular layout. Given the much larger band gap of BP than bilayer graphene [41], we select $V_T$ and $V_B$ values to ensure conduction is limited to bilayer graphene, with BP acting as an effective capacitor and adding strain at the interface (see Figs. S1 and S2 in the Supplemental Material [42]). The device exhibits linear longitudinal current-voltage characteristics (Fig. S3 in the Supplemental Material [42]), indicating good ohmic contacts. The cotunable longitudinal resistance $R$ as a function of carrier density ($n - n_0$) and $D$ [Fig. 1(c)] further demonstrates the $D$-modulated band gap of bilayer graphene [38], where $n_0 \sim 0.1 \times 10^{12}$ cm$^{-2}$ corresponds to minor unintentional doping. Resistance peaks at the charge neutrality point (CNP) with Hall resistance sign changes (Fig. S4 in the Supplemental Material [42]). The estimated electron mobility is $\mu_e \approx 24\,200$ cm$^2$ V$^{-1}$ s$^{-1}$, with quantized Hall plateaus at 10 T (Fig. S4 in the Supplemental Material [42]), indicating the high quality of the sample.

*Contact author: liaozm@pku.edu.cn

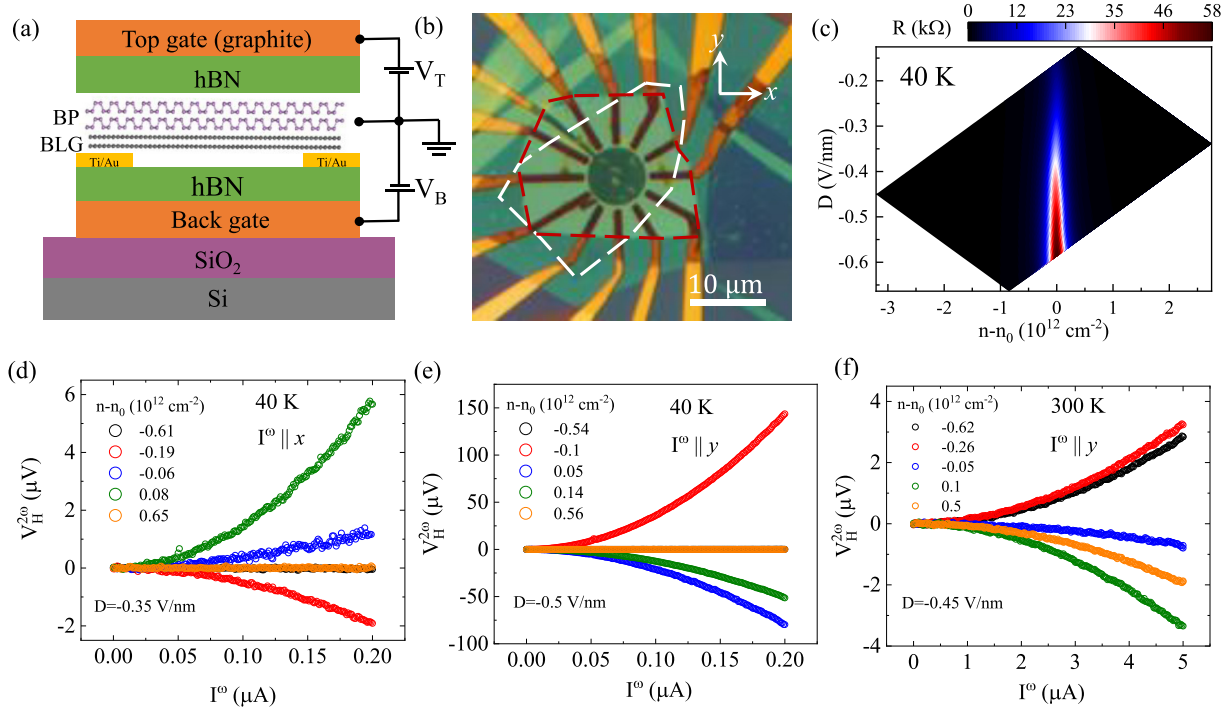

FIG. 1. Nonlinear Hall effect in bilayer graphene/black phosphorus (BP) devices. (a) Schematic of the device. (b) Optical image of the device. BP is marked by the red dotted line, and bilayer graphene is marked by the white dotted line. (c) Longitudinal resistance ($R$) as varying with both carrier density and $D$. Nonlinear Hall voltages $V_H^{2\omega}$ with $I^\omega$ along the (d) $x$ axis and (e) $y$ axis under different carrier densities and a constant $D$ at 40 K. (f) $V_H^{2\omega}$ with $I^\omega$ along the $y$ axis at 300 K.

Due to the symmetry mismatch between bilayer graphene and BP under a perpendicular displacement field, symmetry reduction allows the emergence of the NLHE. When the mirror axes of BP and bilayer graphene are aligned (preserving mirror symmetry $\mathcal{M}_x$), the NLHE is allowed only along the $x$ axis and is zero along the $y$ axis. Misalignment, however, breaks all mirror symmetries, allowing the NLHE along both $x$ and $y$ axes. To measure the NLHE, an AC current ($I^\omega$) with frequency $\omega$ was applied along either the $x$ or $y$ direction, and the second-harmonic Hall voltage ($V_H^{2\omega}$) was measured using a lock-in method. Nonzero $V_H^{2\omega}$ is observed along both $x$ and $y$ axes at different carrier densities under fixed $D$, as shown in Figs. 1(d) and 1(e), respectively. The quadratic dependence of $V_H^{2\omega}$ on $I^\omega$ confirms the second-order nature of the NLHE. The observation of the NLHE for both $I^\omega \parallel x$ and $I^\omega \parallel y$ indicates the absence of mirror symmetry in the misaligned heterostructure. Moreover, the NLHE persists up to room temperature, as shown in Figs. 1(f) and S5 in the Supplemental Material [42]. Control experiments, detailed in Figs. S6 and S7 in the Supplemental Material [42], rule out potential artifacts, including contact effects, crystal anisotropy, and electrode misalignment.

It is worth noting that, for $I^\omega \parallel y$, the NLHE reverses sign only when the carrier type changes from hole to electron [Fig. 1(e)], while for $I^\omega \parallel x$, the NLHE undergoes sign reversal even without a carrier type change, as shown by the red and blue data points in Fig. 1(d). To further understand the NLHE, Figs. 2(a) and 2(b) present $V_H^{2\omega}$ as a function of carrier density and $D$ under a fixed $I^\omega = 100$ nA at 40 K, for $I^\omega \parallel x$ and $I^\omega \parallel y$, respectively. The NLHE is negligible when the Fermi level is far from the CNP but becomes significant near it. Intriguingly, $V_H^{2\omega}$ exhibits overall antisymmetric behavior with respect to electron and hole carriers. For $I^\omega \parallel x$, even within the same carrier type, $V_H^{2\omega}$ demonstrates sign reversal with varying carrier density [Fig. 2(a)]. In contrast, such multiple sign reversals are absent for $I^\omega \parallel y$ [Fig. 2(b)]. Further, Fig. 2(c) highlights that, for $I^\omega \parallel x$, the NLHE reverses sign with changes in the displacement field when the carrier density is fixed near the CNP ($n - n_0 = 0.1 \times 10^{12}$ cm$^{-2}$). However, when $n - n_0 = 0.23 \times 10^{12}$ cm$^{-2}$ (away from the CNP), $V_H^{2\omega}$ does not exhibit sign reversal with varying displacement field.

The phenomenon of multiple sign reversals is observed over a range of temperatures (see Fig. S8 in the Supplemental Material [42]) and occurs at relatively high negative displacement fields ($|D| \geqslant 0.35$ V/nm), indicated by the horizontal dashed line in Fig. 2(a). The critical displacement field required for these reversals $D^*$ shifts to more negative values as temperature increases [Fig. 2(d)]. When $|D| < |D^*|$, multiple sign reversals are likely absent due to charge puddle-induced fluctuations in bilayer graphene [50], which obscure subtle displacement field effects. At higher displacement fields, a larger band gap and more distinct Fermi pockets allow multiple sign changes in the NLHE to appear. As temperature rises, thermal fluctuations further increase $|D^*|$, as shown in Fig. 2(d).

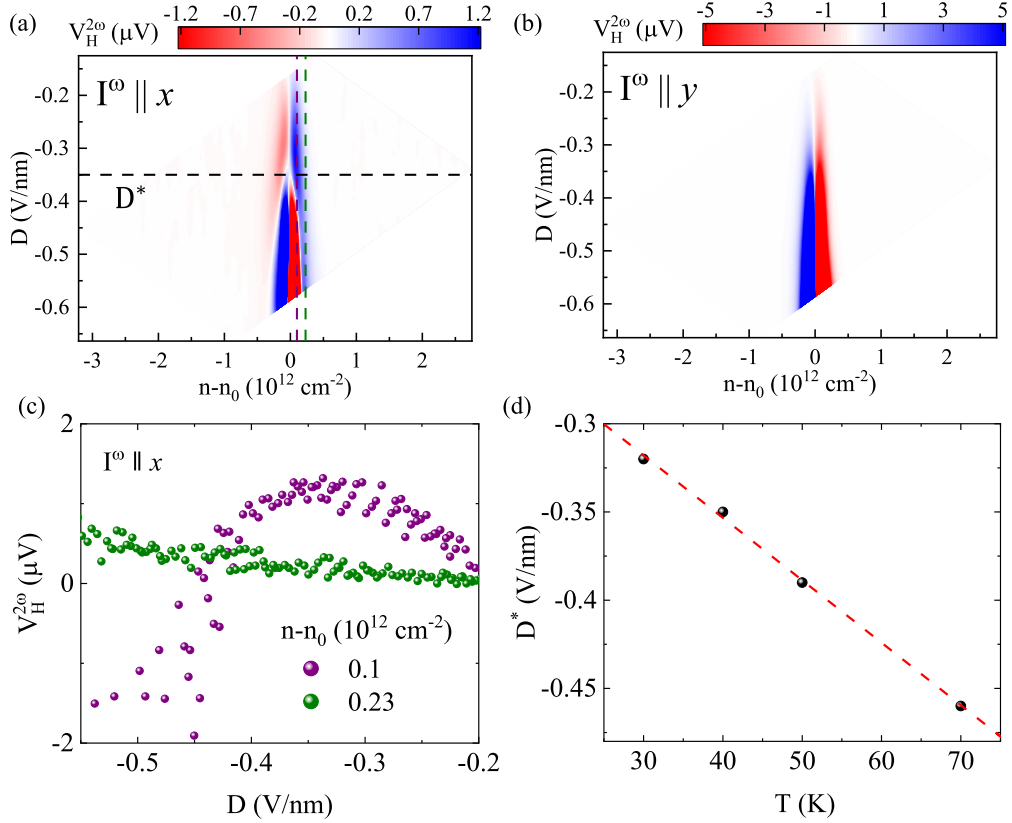

FIG. 2. Gate-switchable nonlinear Hall effect (NLHE) in the bilayer graphene/black phosphorus (BP) heterostructure. $V_H^{2\omega}$ as a function of the carrier density ($n - n_0$) and $D$ under $I^\omega = 100$ nA along the (a) $x$ axis and (b) $y$ axis. (c) $V_H^{2\omega}$ as a function of $D$ under fixed carrier density, corresponding to the vertical line cuts in (a). (d) Critical displacement field $D^*$ as a function of temperature. All data in (a)–(c) were taken at 40 K.

*Scaling analysis.* In addition to the BCD, the NLHE in graphene can arise from skew scattering and side-jump effects, linked to the valley-contrasting chirality of electrons [3]. To separate the contributions from BCD, skew scattering, and side-jump, we conducted a scaling analysis of the NLHE by examining the relationship between NLHE signals and conductivity, varying the displacement field at fixed carrier density, as in Ref. [26]. Generally, a linear scaling law of the NLHE signals is expected: $\frac{V_H^{2\omega}}{(V_L^\omega)^2} = \eta + \xi \sigma_L^2$, where $\frac{V_H^{2\omega}}{(V_L^\omega)^2}$ is the normalized second-harmonic Hall signal, and $\sigma_L^2$ is the square of the longitudinal conductivity $\sigma_L$. Here, the intercept term $\eta$ reflects contributions from the intrinsic BCD and side-jump, while the slope term $\xi \sigma_L^2$ indicates contributions from skew scattering [13]. Figure 3 presents the scaling analysis at 40 K for both $I^\omega \parallel x$ and $I^\omega \parallel y$. The square of the conductivity $\sigma_L^2$ and NLHE signals $\frac{V_H^{2\omega}}{(V_L^\omega)^2}$ are plotted against the displacement field at fixed carrier density in Figs. 3(a) and 3(d). A kink appears in the NLHE signals, as denoted by the blue dotted lines, where the corresponding displacement field is defined as $D_c$. In Figs. 3(b) and 3(e), the relationship between $\frac{V_H^{2\omega}}{(V_L^\omega)^2}$ and $\sigma_L^2$ shows distinct piecewise linear behavior, separated at $D_c$. For $|D| < |D_c|$ (large $\sigma_L$), the linear scaling shows a larger intercept and a smaller slope, indicating dominant contributions from the BCD and side-jump. For $|D| > |D_c|$ (small $\sigma_L$), the slope increases, suggesting enhanced contributions from skew scattering.

To clarify the contributions of the BCD vs side-jump scattering, we analyze the NLHE scaling across different carrier densities, with the intercept of the scaling ($\eta$) shown in Figs. 3(c) and 3(f) as a function of carrier density. It is found that $\eta$ decreases sharply with increasing carrier density, matching the reduction in BCD away from the CNP gap, which suggests the dominated contributions from the BCD. Additionally, by comparing the current directions $I^\omega \parallel x$ and $I^\omega \parallel y$ in Figs. 3(b) and 3(e), we find that, for $I^\omega \parallel y$, the intercept and slope retain their signs across both linear regimes, whereas for $I^\omega \parallel x$, the intercept and slope have opposite signs in the two regions divided by $D_c$. This sign reversal of $\eta$ in Fig. 3(b) reflects a BCD sign change due to varying displacement fields without a carrier-type change.

*Discussions.* To understand the origin of the BCD sign reversals induced by varying displacement fields, we conducted numerical calculations. As reported in Ref. [41], the stacking of graphene and BP can produce moiré superlattices with a pseudo-one-dimensional diagonal stripe pattern. These superlattices create effective lateral strain that breaks symmetry, with the magnitude and orientation of the strain depending on the misalignment angle between BP and graphene. Notably, if the strain is not aligned with the crystalline axes due to misalignment, both the $C_3$ rotational symmetry and mirror symmetry in graphene are broken. Figure 4 shows the calculated Berry curvature and BCD in laterally strained bilayer graphene. Details of the $\mathbf{k} \cdot \mathbf{p}$ Hamiltonian calculations are provided in the Supplemental Material [42]. In bilayer graphene, an applied displacement field enables a tunable band gap, which modifies the Berry curvature and results in a complex Fermi surface topology [47–49]. As depicted in Fig. 4(a), the gapped bilayer graphene host Berry curvature hot spots with opposite signs that emerge at different Fermi pockets. The central Fermi pocket at the $K$ ($K'$) valley exhibits positive (negative) Berry curvature, while the outer Fermi pockets display negative (positive) Berry curvature. Without strain applied to bilayer graphene, the $C_3$ symmetry forbids the appearance of BCD.

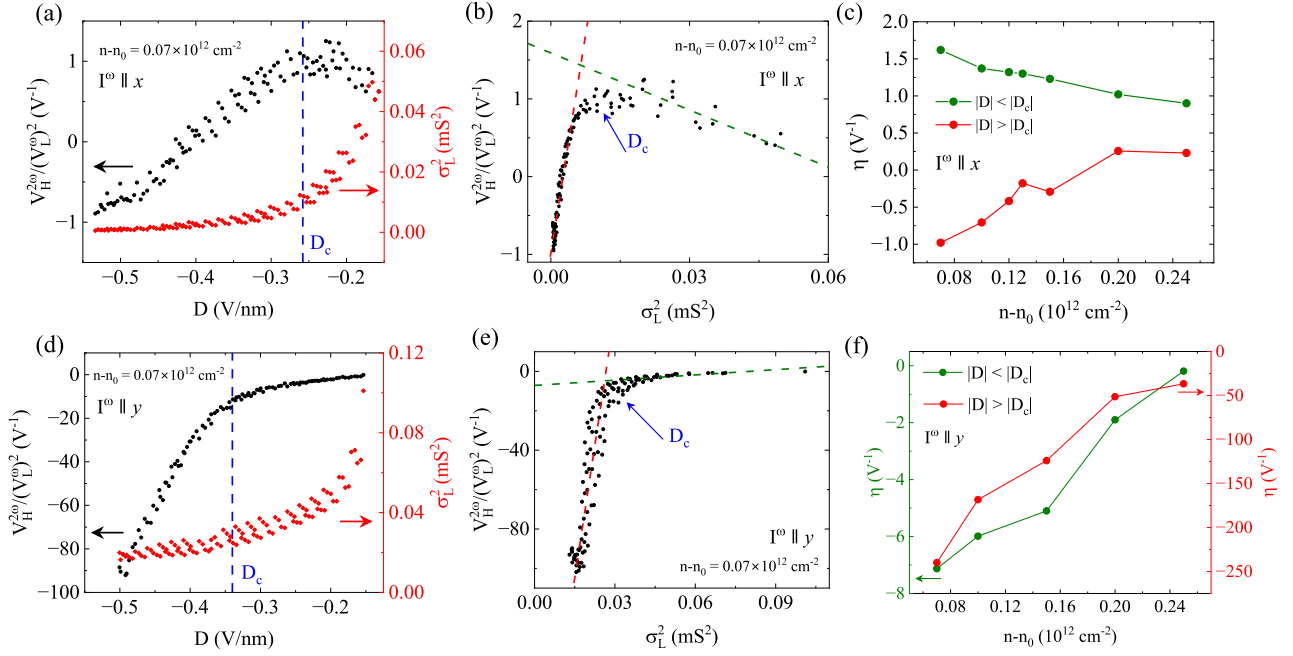

FIG. 3. Scaling analysis of the nonlinear Hall effect (NLHE) at 40 K. NLHE signals $\frac{V_H^{2\omega}}{(V_L^\omega)^2}$ and $\sigma_L^2$ as a function of $D$ at $n - n_0 = 0.07 \times 10^{12}$ cm$^{-2}$ for (a) $I^\omega \parallel x$ and (d) $I^\omega \parallel y$. $\frac{V_H^{2\omega}}{(V_L^\omega)^2}$ as a function of $\sigma_L^2$ at $n - n_0 = 0.07 \times 10^{12}$ cm$^{-2}$ for (b) $I^\omega \parallel x$ and (e) $I^\omega \parallel y$. The data points in (b) and (e) are collected from (a) and (d), respectively. Intercept of the linear scaling $\eta$ as a function of carrier density for (c) $I^\omega \parallel x$ and (f) $I^\omega \parallel y$.

After stacking with BP, the moiré superlattice induces effective lateral strain in bilayer graphene, resulting in $C_3$ symmetry breaking [41]. The symmetry breaking is shown in Fig. 4(b), where the three outer Berry curvature hotspots are no longer equivalent. Further, the mirror symmetry $\mathcal{M}_x$ is also broken since the lateral strain deviates from the crystalline axes. These symmetry breakings in bilayer graphene lead to a BCD along both $x$ and $y$ axes, which therefore contribute to the NLHE. Figures 4(d) and 4(f) show the BCD components $\Lambda_x$ and $\Lambda_y$ along $x$ and $y$ axes, respectively. Notably, $\Lambda_x$ shows three sign changes, where the sign reversal can occur without the change of carrier type at both hole and electron sides [Fig. 4(d)]. By comparison, such multiple sign-change behavior is absent for $\Lambda_y$ [Fig. 4(f)]. The calculation results reproduce the similar evolution of $V_H^{2\omega}$ concerning carrier density, as shown in Figs. 4(c) and 4(e) for $I^\omega \parallel x$ and $I^\omega \parallel y$, respectively.

Further details on the effects of lateral strain are provided in Figs. S9 and S10 in the Supplemental Material [42]. It is found that the multiple sign reversals of the BCD are robust against varying the strain directions and magnitudes. A strain effect term $|w| \approx 2$ meV, corresponding to ~0.34% strain in bilayer graphene [49], is sufficient to induce these sign reversals. Considering the 3–5% strain typically observed in graphene/BP [41], the alignment between our theoretical calculations and experimental data suggests that the multiple sign reversals of the BCD primarily originate from effective lateral strains in the bilayer graphene/BP stacking.

In summary, we demonstrate the generation and modulation of the NLHE up to room temperature in bilayer graphene/BP heterostructures. By stacking graphene with BP, effective lateral strain breaks the symmetry of the system, inducing a nonzero BCD and the NLHE. Scaling analysis shows that BCD, skew scattering, and side jump all contribute to the NLHE. Combined with numerical calculations, our findings reveal that the BCD can be precisely tuned via carrier density and displacement field. Notably, the nonlinear Hall voltage undergoes a sign reversal as carrier density changes, even for carriers of the same type, which correlates with the polarity change of the BCD as the Fermi level shifts. This reflects unique fermiology, with Berry curvature distributions differing across Fermi pockets in bilayer graphene. In this letter, we advance symmetry engineering in 2D materials through van der Waals stacking [51–54], supporting the development of room-temperature quantum nonlinear devices in graphene-based systems.

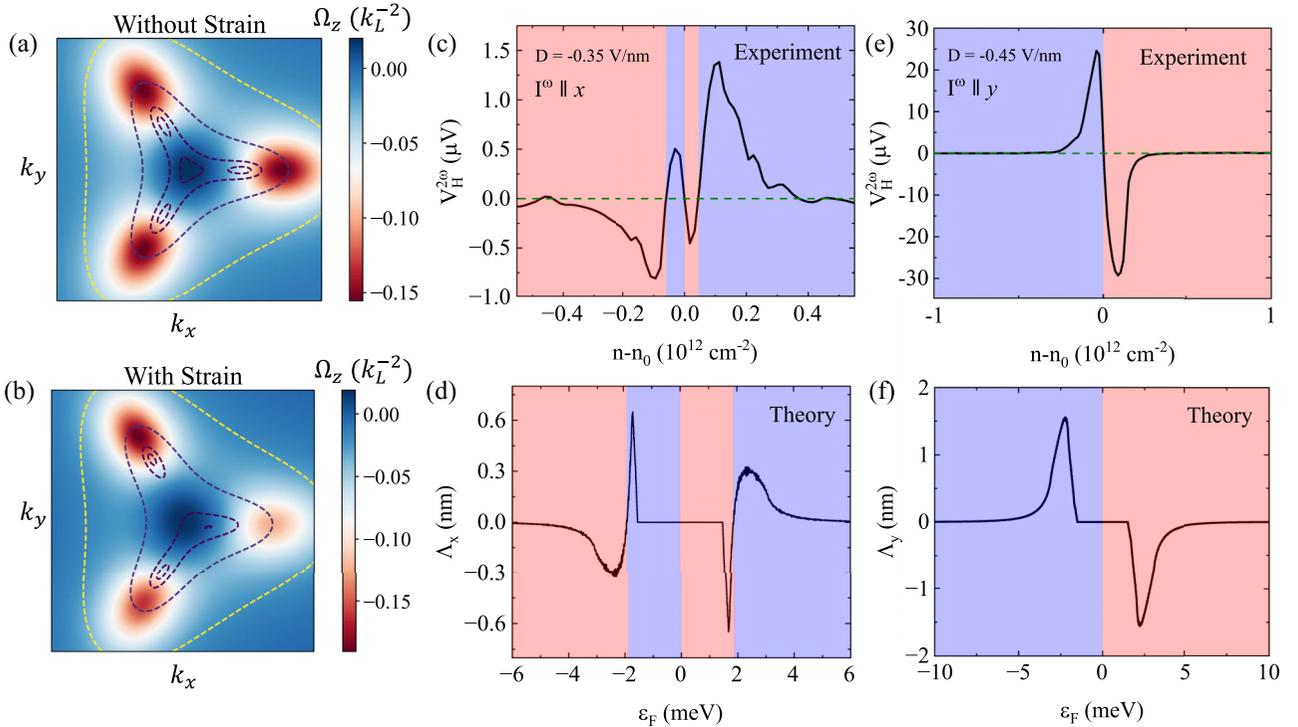

FIG. 4. Strain-induced symmetry breaking and nonzero Berry curvature dipole (BCD) in bilayer graphene. Berry curvature of bilayer graphene at the $K$ valley of the conductance band (a) without and (b) with lateral strains, respectively. Dotted lines represent the constant-energy contours. Berry curvature $\Omega_z$ is in units of $\kappa_L^{-2}$ with $\kappa_L \approx 0.035$ nm$^{-1}$. Experimentally measured $V_H^{2\omega}$ with $I^\omega$ along the (c) $x$ axis and (e) $y$ axis as a function of carrier density under a constant $D$. Theoretical calculation of the BCD: (d) $\Lambda_x$ and (f) $\Lambda_y$ vs chemical potential in lateral strained bilayer graphene with a 3 meV band gap.

*Acknowledgments.* This letter was supported by the National Natural Science Foundation of China (Grants No. 62425401 and No. 62321004) and the Innovation Program for Quantum Science and Technology (Grant No. 2021ZD0302403).

# Supplemental Materials for:

## Engineering Nonlinear Hall Effect in Bilayer Graphene/Black Phosphorus Heterostructure

Xing-Guo Ye[1], Zhen-Tao Zhang[1], Peng-Fei Zhu[1], Wen-Zheng Xu[1], An-Qi Wang[1], Zhi-Min Liao[1,2*]

[1]State Key Laboratory for Mesoscopic Physics and Frontiers Science Center for Nano-optoelectronics, School of Physics, Peking University, Beijing 100871, China.
[2]Hefei National Laboratory, Hefei 230088, China.

*Corresponding author. Email: liaozm@pku.edu.cn

**Methods**

**Device Fabrication.** Employing the standard mechanical exfoliation method, we obtained bilayer graphene and few-layer black phosphorus flakes from the bulk crystals (from HQ Graphene). The hBN/graphite was transferred onto a $SiO_2$/Si substrate to sever as the bottom gate using a polymer-based dry transfer technique. Ti/Au electrodes were then patterned on this hBN layer, and the residual adhesive was cleaned up by atomic force microscope. The top-gate graphite, hBN, black phosphorus, and bilayer graphene were successively picked up and transferred onto the pre-cleaned electrodes using a dry transfer technique [43]. The entire exfoliation and transfer process took place in an argon-filled glove box with $O_2$ and $H_2O$ content maintained below 0.01 parts per million to prevent sample degeneration.

**Transport Measurements.** Transport measurements were carried out in an Oxford cryostat with a variable temperature insert and a superconducting magnet. First- and second-harmonic signals were collected by standard lock-in techniques (Stanford Research Systems Model SR830) with frequency $\omega$. Frequency $\omega$ equals 17.777 Hz unless otherwise stated. The gate voltage was applied through Keithley 2400 SourceMeters.

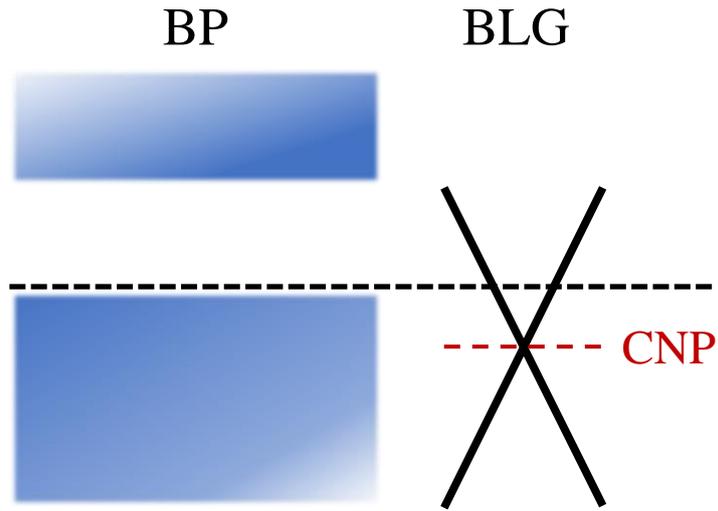

**Fig. S1. Band alignment between bilayer graphene (BLG) and black phosphorus (BP).**

Besides effective strain, the interactions at the bilayer graphene/black phosphorus heterointerface are complex, encompassing band mixing, moiré potential, band folding, and other effects. These factors may also contribute to the nonlinear Hall effect and its multiple sign changes in this system. However, modeling few-layer black phosphorus as an effective capacitor and introducing strain in bilayer graphene effectively explains the experimental observations for the following reasons:

(1) **Work Function Considerations:** The work functions of bilayer graphene and black phosphorus are 4.69 eV and 4.47 eV, respectively [41,44]. Thus, the charge neutral point of bilayer graphene is closer to the valence band edge of black phosphorus. As shown in Fig. S1, the Fermi level is close to the hole side of black phosphorus. In our measurements, the top gate voltage $V_T$ is set to be larger than 1 V, ensuring the Fermi level lies in the bandgap of black phosphorus. Moreover, as discussed in Fig. S2, the effective gate capacitance is accurately estimated only when considering black phosphorus, further confirming its role as an effective capacitor.

**(2) Electron-Hole Symmetry:** As shown in Fig. S1, band mixing and band folding primarily occur on the hole side of bilayer graphene due to the work function misalignment between black phosphorus and bilayer graphene, which induces electron-hole asymmetry. If the multiple sign changes of the nonlinear Hall effect were primarily caused by band mixing or band folding, the phenomenon should be more significant on the hole side and suppressed on the electron side. However, as shown in Fig. 2(a), the results are almost symmetric on both electron and hole sides. This observation is inconsistent with the band mixing scenario but rather supports the strained bilayer graphene model.

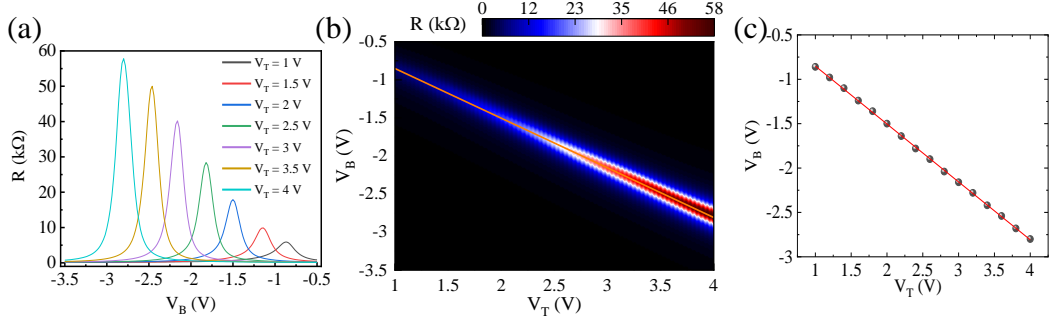

**Fig. S2. Longitudinal resistance as a function of gate voltages.**

**a,** Longitudinal resistance plotted against back gate voltage at different top gate voltages. Increasing top gate voltage results in higher resistance peaks, indicative of an enlarged energy gap in bilayer graphene.

**b,** Mapping of longitudinal resistance at T = 40 K; the orange line denotes the peak resistance.

**c,** Relationship between back and top gate voltages at the resistance peak position.

Dual gates ($V_T, V_B$) enable the independent tuning of carrier density $n$ and displacement field $D$, following by $n = (C_B V_B + C_T V_T)/e$, and $D = (C_B V_B - C_T V_T)/2\varepsilon_0$, where $\varepsilon_0 = 8.85 \times 10^{-12}$ Fm$^{-1}$ is the vacuum permittivity, $e$ is the electron charge, $C_T$ and $C_B$ are the top and back gate capacitance, respectively. $C_B = \frac{\varepsilon_0 \varepsilon_{BN}}{d_{bBN}}$ is the capacitance of bottom hBN, where $\varepsilon_{BN} \approx 3$ is the relative dielectric constant of the hBN layers, and $d_{bBN} = 18$ nm is the thickness of bottom hBN. $C_T = (\frac{d_{tBN}}{\varepsilon_0 \varepsilon_{BN}} + \frac{d_{BP}}{\varepsilon_0 \varepsilon_{BP}})^{-1}$, where $d_{tBN} = 22$ nm is the thickness of top hBN, $d_{BP} = 12$ nm is the thickness of black phosphorus, and $\varepsilon_{BP} \approx 6.55$ is the relative dielectric constant of black phosphorus. To confirm that black phosphorus does not participate in electric conduction here and acts as an effective capacitor, we select the range of ($V_T, V_B$) values where the charge-neutral point satisfying $n - n_0 = (C_B V_B + C_T V_T)/e - n_0 = 0$, as shown in Fig. S2(c). In this selected range of ($V_T, V_B$) values, the $V_T$ vs. $V_B$ exhibits a linear relationship with a slope of -0.652, closely aligning with the theoretically calculated slope $\alpha = -\frac{C_T}{C_B} \approx -0.655$ when considering black phosphorus as a capacitor.

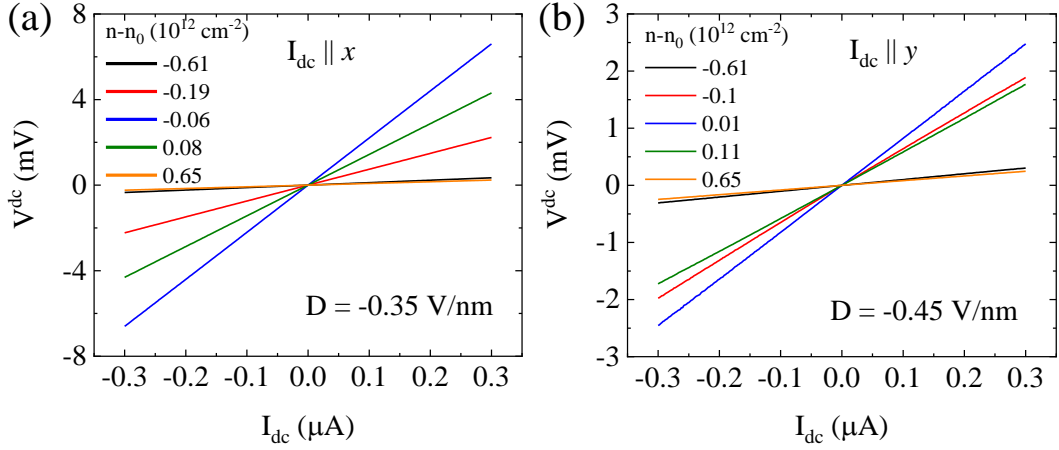

**Fig. S3. The voltage-current relationship with current along the (a) *x*-axis and (b) *y*-axis at 40 K under varying carrier density and a fixed displacement field *D*.**

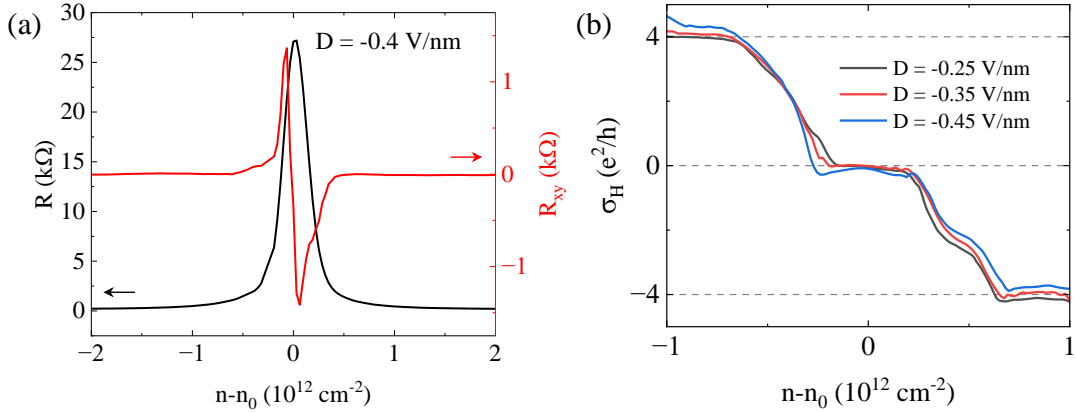

**Fig. S4. The Hall effect in the bilayer graphene/black phosphorus heterostructure at 1.6 K.**

**a,** The longitudinal resistance and Hall resistance as a function of carrier density under $D = -0.4$ V/nm at magnetic field $B = 0.02$ T. The electron mobility is estimated about $\mu_e \approx 24200 \ cm^2 \cdot V^{-1} \cdot s^{-1}$.

**b,** The quantized Hall conductance at magnetic field $B = 10$ T.

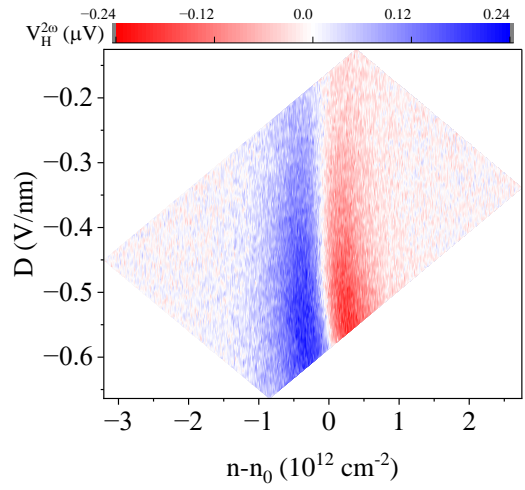

**Fig. S5.** The nonlinear Hall voltages as a function of displacement field and carrier density under $I^\omega = 1$ μA at 300 K.

**Extrinsic second-order responses**

In addition to intrinsic effects such as the Berry curvature dipole and scattering mechanisms, other possible extrinsic effects could also contribute to the second-order transverse response. Here, through control experiments, these extrinsic effects have been carefully ruled out one by one.

1. *Diode effect.*

Contact-induced diodes are typically unavoidable between metal electrodes and two-dimensional materials, often leading to higher-order transport effects. However, two-terminal I-V measurements (Fig. S3) show linear behavior, ruling out the extrinsic diode effect as the source of the nonlinear Hall effect.

2. *Mixing from the longitudinal transport.*

The transverse transport can be induced by the mixing from the longitudinal transport when there is electrodes misalignment, charge inhomogeneity, or crystalline anisotropy [45]. In all these cases, since the induced electric field is no longer parallel to the current, a significant transverse voltage is expected.

The first-order longitudinal ($V_L^\omega$) and transverse ($V_H^\omega$) voltage for $I^\omega \parallel x$ and $I^\omega \parallel y$ are measured as shown in Fig. S6. Clearly, the transverse voltage $V_H^\omega$ remains small, which is only ~1% of $V_L^\omega$ for both $I^\omega \parallel x$ and $I^\omega \parallel y$. If this transverse voltage is induced by electrodes misalignment, the small $V_H^\omega$ suggests the misalignment angle is less than 1°. This observation shows that the current applied in the bilayer graphene has a negligible transverse component, ruling out extrinsic factors like electrode misalignment and resistance anisotropy as primary causes of the observed nonlinear Hall effect.

Further, both the second-harmonic longitudinal ($V_L^{2\omega}$) and Hall ($V_H^{2\omega}$) voltage for $I^\omega \parallel x$ and $I^\omega \parallel y$ are shown in Fig. S7. Clearly, the nonlinear Hall voltages $V_H^{2\omega}$ dominate over the nonlinear longitudinal voltages $V_L^{2\omega}$ for both $I^\omega \parallel x$ and $I^\omega \parallel y$. This observation also excludes the transverse component of the nonlinear longitudinal signals as the origin of the observed nonlinear Hall signals.

### 3. *Thermal effect and thermoelectric effect*

The Joule heating can modify the sample resistance, thus leading to the nonlinear transport [46]. Note the Joule heating is expected to be more significant when the resistance is larger. Therefore, the thermal-induced nonlinear voltages are expected to be more significant for $I^\omega \parallel x$ due to the larger resistance along the *x*-axis (e.g. Fig. S3). However, the observed nonlinear Hall voltage shows a more significant value for $I^\omega \parallel y$ (Figs. 2(a)-(b)). Such inconsistence clearly rules out the thermal-related effect as the main origin of the nonlinear Hall effect.

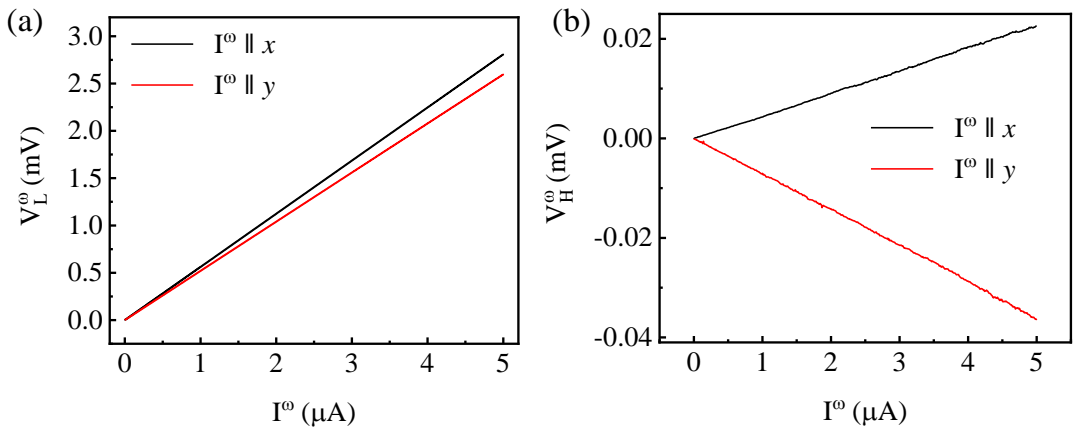

**Fig. S6. The first-order (a) longitudinal ($V_L^\omega$) and (b) transverse ($V_H^\omega$) voltage for $I^\omega \parallel x$ and $I^\omega \parallel y$, respectively, at 1.6 K.**

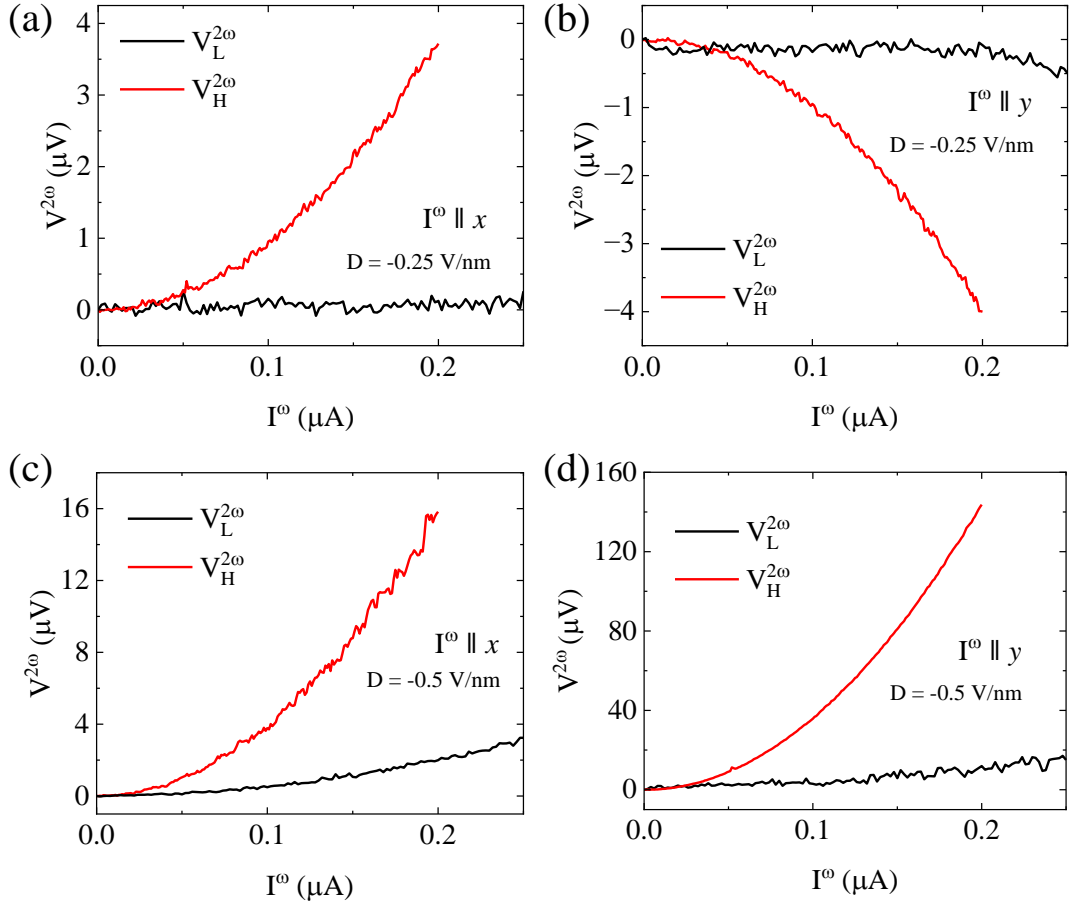

**Fig. S7. Comparison between longitudinal and Hall nonlinear signals.**

**a,b,** The second-harmonic longitudinal ($V_L^{2\omega}$) and Hall ($V_H^{2\omega}$) voltage for (**a**) $I^\omega \parallel x$ and (**b**) $I^\omega \parallel y$ at displacement field D = -0.25 V/nm and the carrier density $n - n_0 = 0.04 \times 10^{12}$ cm$^{-2}$.

**c,d,** The second-harmonic longitudinal ($V_L^{2\omega}$) and Hall ($V_H^{2\omega}$) voltage for (**c**) $I^\omega \parallel x$ and (**d**) $I^\omega \parallel y$ at displacement field D = -0.5 V/nm and the carrier density $n - n_0 = -0.1 \times 10^{12}$ cm$^{-2}$.

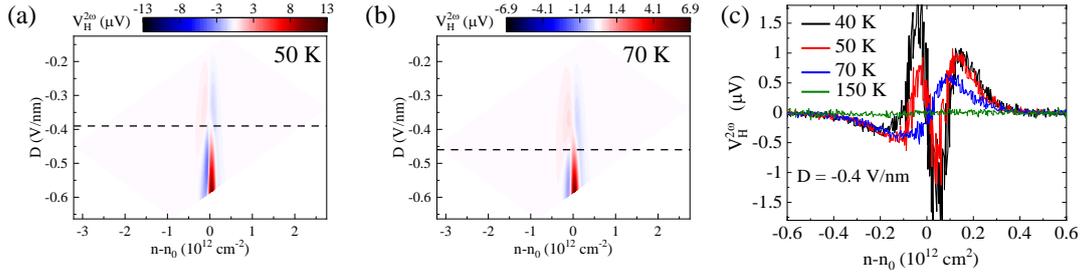

**Fig. S8. Temperature dependence of the nonlinear Hall effect.**

**a,b,** 2D map of nonlinear Hall effect at (**a**) 50 K and (**b**) 70 K, respectively.

**c,** The relationship between $V_H^{2\omega}$ and carrier density measured at $D = -0.4$ V/nm and $I^{\omega} = 100$ nA at different temperatures.

As shown in Fig. S8, multiple sign changes of the nonlinear Hall effect can indeed be observed at higher temperatures by increasing the displacement fields. It is found that the critical electric displacement field D*, which is the onset displacement field at which multiple sign changes occur, increases with temperature (Fig. 2(d)). The maximum displacement field accessible in our work is approximately -0.6 V/nm. Within this range, multiple sign changes of the nonlinear Hall effect can be observed at temperatures up to 70 K, as shown in Fig. S8(b). However, in Fig. S8(c), the displacement field is D = −0.4 V/nm, which is smaller than the critical field D* = −0.46 V/nm at 70 K. Consequently, multiple sign changes are not observed for D = −0.4 V/nm at 70 K in Fig. S8(c).

**Lateral strained bilayer graphene**

The effective Hamiltonian calculations for bilayer graphene are based on the methodology outlined in Ref. [47-49]. The resulting $2 \times 2$ Hamiltonian, derived from the low-energy effective $\mathbf{k} \cdot \mathbf{p}$ Hamiltonian, is given by:

$$H = \left(-\frac{1}{2m}(k_x^2 - k_y^2) + \xi v_3 k_x + w\right)\sigma_x - \left(\frac{1}{m}k_x k_y + \xi v_3 k_y\right)\sigma_y + \frac{\Delta}{2}\sigma_z,$$

where the characteristic energy $\varepsilon_L = \frac{1}{2}mv_3^2 \approx 1$ meV, the characteristic momentum $\kappa_L = mv_3 \approx 0.035$ nm$^{-1}$, $\sigma$ is the Pauli matrix, $\Delta$ is the bandgap, $\xi = \pm 1$ denotes the valley index. As shown in Ref.[41], effective lateral strain can be induced in the black phosphorus/graphene heterostructures, where the magnitude and orientation of the strain vary with the misalignment angle between black phosphorus and graphene. We include the effective lateral strain by the strain tensor $u_{\alpha\beta} = \frac{1}{2}(\partial_\alpha \delta r_\beta + \partial_\beta \delta r_\alpha)$ ($\alpha, \beta = x, y$ and $\boldsymbol{\delta r} = (\delta r_x, \delta r_y)$ stands for the displacement). The strain tensor can be further described by its eigenvalues $\delta$ and $\delta'$, and the angle $\theta$ between the principal axis of strain tensor and the $x$-axis (Fig. S9(a)), where $\theta = 0°$ corresponds to the mirror axes alignment between black phosphorus and bilayer graphene. The lateral strain is described by the strain term $w$ in the Hamiltonian of bilayer graphene with $w \propto (\delta - \delta')e^{-2i\theta}$ [49]. Here we ignore the interlayer shear. The Berry curvature dipole is calculated according to $\Lambda_\alpha = \frac{1}{\hbar}\int f_0(k, \varepsilon_F)\frac{\partial \Omega_z}{\partial k_\alpha}d^2\boldsymbol{k}$, where $f_0(k, \varepsilon_F)$ is the equilibrium distribution function.

Figure S9 shows the Berry curvature dipole as a function of Fermi energy at various strain directions $\theta$. Clearly, the multiple sign-reversal of Berry curvature dipole along $x$-axis ($\Lambda_x$) is robust against the strain direction $\theta$. By comparison, such multiple sign-reversal is absent for $\Lambda_y$. Moreover, when $\theta = 0°$, the presence of the mirror symmetry forces the $\Lambda_y = 0$. When $\theta$ is increased, $\Lambda_y$ is rapidly increased with $\Lambda_y > \Lambda_x$ even at small $\theta$.

Figure S10 shows the Berry curvature dipole as a function of Fermi energy at different strain magnitudes $|w|$. The multiple sign-reversal of the $\Lambda_x$ is observed at all the strain magnitudes, indicating the robustness against $|w|$. By comparison, such multiple sign-reversal is absent for $\Lambda_y$ at different strain magnitudes.

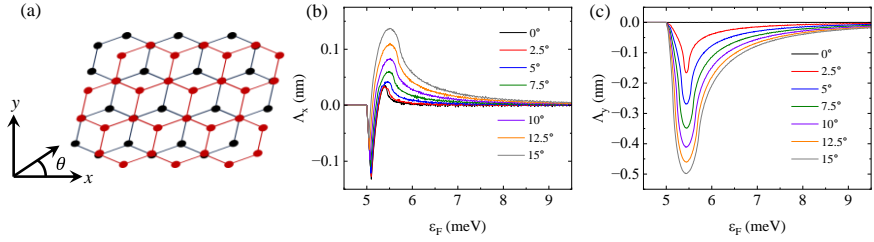

**Fig. S9. Strain direction-dependence of the Berry curvature dipole.**

**a,** The bilayer graphene under lateral strain with the principal direction $\theta$.

**b,c,** The Berry curvature dipole (b) along *x* axis and (c) along *y* axis as a function of Fermi energy at various strain principal directions $\theta$ with $|w| \approx 2$ meV.

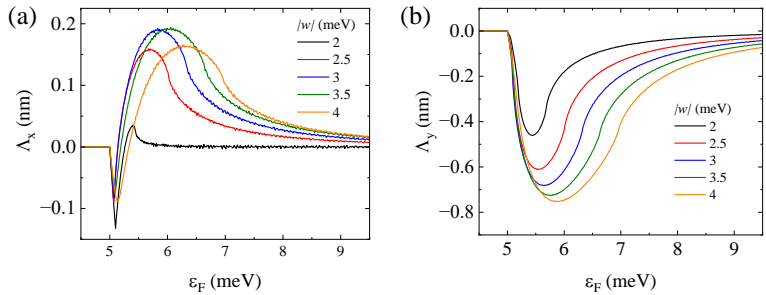

**Fig. S10. Strain magnitude-dependence of the Berry curvature dipole.**

**a,b,** The Berry curvature dipole (a) along *x* axis and (b) along *y* axis as a function of Fermi energy at various strain magnitudes $|w|$.